\documentclass{article}
\usepackage{graphicx}
\usepackage{amsmath}

\input{tcilatex}

\begin{document}

\title{\textbf{Gravitation with Two Times}}
\author{W. Chagas-Filho \\
Departamento de Fisica, Universidade Federal de Sergipe\\
SE, Brazil}
\maketitle

\begin{abstract}
We investigate the possibility of constructing a covariant Newtonian
gravitational theory and find that the action describing a massless
relativistic particle in a background Newtonian gravitodynamic field has a
higher-dimensional extension with two times..
\end{abstract}

\noindent

\section{Introduction}

There is at present time a considerable amount of effort directed to the
construction of a fully consistent quantum theory for the gravitational
interaction. Some time ago such a theory was expected to be given by string
theory [1]. String theory instead revealed itself as having many dual
formulations and this fact was interpreted as an indication that these dual
string formulations, together with supergravity, were only different aspects
of a more fundamental theory in a higher dimension, which is now called
M-theory [2,3]. A consistent quantum theory for gravity was then expected to
be given by M-theory. This theory had a considerable success in dealing with
gravitational physics. In a particular formulation the quantum aspects of
gravity are reproduced, in the low and high energy limits, by the collective
behavior of the Dirichlet $p=0$ branes, also called D-particles, that
compose the membrane structure which underlies this formulation of M-theory.
Formulated in terms of D-particles, the quantum mechanics of M-theory can
describe consistently both the Newtonian (low energy limit) and the general
relativistic (high energy limit) aspects of gravity [2,3].

The purpose of this paper is to call the attention of researchers in quantum
gravity and related topics to the fact that in the gap between Newtonian
gravitation and Einsteinian gravitation there may exist another formulation
of gravity, described by a relativistic Newtonian gravitational theory. This
relativistic gravitational theory can be naturally constructed after
introducing a gravitomagnetic field [4], the gravitational analogue of the
magnetic field, in Newton's original theory for the gravitational
interaction. Because a consistent quantum theory for gravity is still
lacking, and the spin 2 gauge particle predicted by general relativity has
not yet been detected, there is in principle nothing wrong in trying to
investigate this relativistic Newtonian gravitational theory. In practical
terms a bad conclusion of this investigation could be, for instance, that
gravitodynamics is as observable as supersymmetry, and that only it's low
energy limit (Newtonian gravitation) can be observable. However, as
supersymmetry, it may have interesting theoretical applications and may even
lead to new insights in theoretical physics. One of these theoretical
applications would be the study of quantum gravity at intermediate energy
scales, in an attempt to construct a bridge between the information about
gravity available from the quantum behaviour of M-theory described above.

In this paper we try some steps in the investigation of this relativistic
Newtonian gravitational theory and arrive at surprising conclusions. We
study a simple model defined by the action describing a massless
relativistic particle in a background gravitodynamic potential. This action
defines a reparametrization invariant system. It has a $d$ dimensional
conformal invariance. As is well known [5,6,7,8,9,10], the $d$ dimensional
conformal invariance of massless particle systems is identical to the
Lorentz invariance in a $d+2$ dimensional space-time. This indicates that
our model gravitodynamic action has a higher-dimensional extension. The
Hamiltonian corresponding to this action has a scale invariance that can be
used to induce the appearance in phase space of the classical analogues of
the Snyder commutators [11]. The action naturally fits into the formalism of
two-time physics [5,6,7,8,9].

The article is organized as follows. In section two we review the
invariances of the free massless relativistic particle action in a $d$
dimensional Euclidean space-time and show how a scale invariance of the
Hamiltonian induces Snyder-like brackets in the particle's phase space. In
section three we perform the same analysis in the presence of a background
gravitodynamic field. In section four we extend our gravitodynamic action to
an action in $d+2$ dimensions. We also show how, in a particular gauge, this 
$d+2$ dimensional theory reproduces the conformal invariance of the $d$
dimensional action of section three. Some concluding remarks appear in
section five.

\section{Massless Particles}

A free massless relativistic particle in a $d$ dimensional flat Euclidean
space-time is described by the action 
\begin{equation}
S=\frac{1}{2}\int d\tau \lambda ^{-1}\dot{x}^{2}  \tag{2.1}
\end{equation}
Action (2.1) is invariant under the Poincar\'{e} transformation 
\begin{equation}
\delta x^{\mu }=a^{\mu }+\omega _{\nu }^{\mu }x^{\nu }  \tag{2.2a}
\end{equation}
\begin{equation}
\delta \lambda =0  \tag{2.2b}
\end{equation}
under global scale transformations 
\begin{equation}
\delta x^{\mu }=\alpha x^{\mu }  \tag{2.3a}
\end{equation}
\begin{equation}
\delta \lambda =2\alpha \lambda  \tag{2.3b}
\end{equation}
where $\alpha $ is a constant, and under the conformal transformations 
\begin{equation}
\delta x^{\mu }=(2x^{\mu }x^{\nu }-\delta ^{\mu \nu }x^{2})b_{\nu } 
\tag{2.4a}
\end{equation}
\begin{equation}
\delta \lambda =4\lambda x.b  \tag{2.4b}
\end{equation}
where $b_{\mu }$ is a constant vector. As a consequence of the presence of
these invariances we can define in space-time the following field 
\begin{equation}
V=a^{\mu }p_{\mu }-\frac{1}{2}\omega ^{\mu \nu }M_{\mu \nu }+\alpha D+b^{\mu
}K_{\mu }  \tag{2.5}
\end{equation}
with the generators 
\begin{equation}
p_{\mu }  \tag{2.6a}
\end{equation}
\begin{equation}
M_{\mu \nu }=x_{\mu }p_{\nu }-x_{\nu }p_{\mu }  \tag{2.6b}
\end{equation}
\begin{equation}
D=x.p  \tag{2.6c}
\end{equation}
\begin{equation}
K_{\mu }=2x_{\mu }x.p-x^{2}p_{\mu }  \tag{2.6d}
\end{equation}
$p_{\mu }$ generates translations in space-time,$\ M_{\mu \nu }$ generates
space-time rotations, $D$ is the generator of space-time dilatations and $%
K_{\mu }$ generates conformal transformations. These generators define the
algebra 
\begin{equation}
\{p_{\mu },p_{\nu }\}=0  \tag{2.7a}
\end{equation}
\begin{equation}
\{p_{\mu },M_{\nu \lambda }\}=\delta _{\mu \nu }p_{\lambda }-\delta _{\mu
\lambda }p_{\nu }  \tag{2.7b}
\end{equation}
\begin{equation}
\{M_{\mu \nu },M_{\lambda \rho }\}=\delta _{\nu \lambda }M_{\mu \rho
}+\delta _{\mu \rho }M_{\nu \lambda }-\delta _{\nu \rho }M_{\mu \lambda
}-\delta _{\mu \lambda }M_{\nu \rho }  \tag{2.7c}
\end{equation}
\begin{equation}
\{D,D\}=0  \tag{2.7d}
\end{equation}
\begin{equation}
\{D,p_{\mu }\}=p_{\mu }  \tag{2.7e}
\end{equation}
\begin{equation}
\{D,M_{\mu \nu }\}=0  \tag{2.7f}
\end{equation}
\begin{equation}
\{D,K_{\mu }\}=-K_{\mu }  \tag{2.7g}
\end{equation}
\begin{equation}
\{p_{\mu }.K_{\nu }\}=-2\delta _{\mu \nu }D+2M_{\mu \nu }  \tag{2.7h}
\end{equation}
\begin{equation}
\{M_{\mu \nu },K_{\lambda }\}=\delta _{\nu \lambda }K_{\mu }-\delta
_{\lambda \mu }K_{\nu }  \tag{2.7i}
\end{equation}
\begin{equation}
\{K_{\mu },K_{\nu }\}=0  \tag{2.7j}
\end{equation}
computed in terms of the Poisson brackets 
\begin{equation}
\{p_{\mu },p_{\nu }\}=0  \tag{2.8a}
\end{equation}
\begin{equation}
\{x_{\mu },p_{\nu }\}=\delta _{\mu \nu }  \tag{2.8b}
\end{equation}
\begin{equation}
\{x_{\mu },x_{\nu }\}=0  \tag{2.8c}
\end{equation}
The algebra (2.7) is the conformal space-time algebra [10]. The free
massless particle theory defined by action (2.1) is a conformal theory in $d$
dimensions. The important point here is that the algebra (2.7) is also the
Lorentz algebra in $d+2$ dimensions

The classical equation of motion for $x^{\mu }$ that follows from action
(2.1) is 
\begin{equation*}
\frac{d}{d\tau }(\frac{\dot{x}_{\mu }}{\lambda })=0
\end{equation*}
The equation of motion for $\lambda $ gives the condition $\dot{x}^{2}=0$,
which tells us that a free massless relativistic particle moves at the speed
of light. As a consequence of this, it becomes impossible to solve for $%
\lambda (\tau )$ from its equation of motion. In the massless theory the
value of $\lambda (\tau )$ is arbitrary.

In the transition to the Hamiltonian formalism action (2.1) gives the
canonical momenta 
\begin{equation}
p_{\lambda }=0  \tag{2.9}
\end{equation}
\begin{equation}
p_{\mu }=\frac{\dot{x}_{\mu }}{\lambda }  \tag{2.10}
\end{equation}
and the canonical Hamiltonian 
\begin{equation}
H=\frac{1}{2}\lambda p^{2}  \tag{2.11}
\end{equation}
Equation (2.9) is a primary constraint. Introducing the Lagrange multiplier $%
\xi (\tau )$ for this constraint we can write the Dirac Hamiltonian 
\begin{equation}
H_{D}=\frac{1}{2}\lambda p^{2}+\xi p_{\lambda }  \tag{2.12}
\end{equation}
Requiring the dynamical stability of constraint (2.9), $\dot{p}_{\lambda
}=\{p_{\lambda },H_{D}\}=0$, we obtain the secondary constraint 
\begin{equation}
\phi =\frac{1}{2}p^{2}=0  \tag{2.13}
\end{equation}
Constraint (2.13) has a vanishing Poisson bracket with constraint (2.9),
being therefore a first-class constraint [12]. Constraint (2.9) generates
translations in the arbitrary variable $\lambda (\tau )$ and can be dropped
from the formalism.

The massless particle Hamiltonian (2.11) is invariant under the
transformations 
\begin{equation}
p_{\mu }\rightarrow \tilde{p}_{\mu }=\exp \{-\beta \}p_{\mu }  \tag{2.14a}
\end{equation}
\begin{equation}
\lambda \rightarrow \exp \{2\beta \}\lambda  \tag{2.14b}
\end{equation}
where $\beta $ is an arbitrary function of $x$ and $p$. From the equation
(2.10) for the canonical momentum we find that $x^{\mu }$ should transform
as 
\begin{equation}
x^{\mu }\rightarrow \tilde{x}^{\mu }=\exp \{\beta \}x^{\mu }  \tag{2.14c}
\end{equation}
when $p_{\mu }$ transforms as in (2.14a).

Consider now the bracket structure that transformations (2.27a) and (2.28)
induce in the massless particle phase space. Retaining only the linear terms
in $\beta $ in the exponentials, we find that the new transformed canonical
variables $(\tilde{x}_{\mu },\tilde{p}_{\mu })$ obey the brackets 
\begin{equation}
\{\tilde{p}_{\mu },\tilde{p}_{\nu }\}=0  \tag{2.15a}
\end{equation}
\begin{equation}
\{\tilde{x}_{\mu },\tilde{p}_{\nu }\}=(1+\beta )[\delta _{\mu \nu }(1-\beta
)-\{x_{\mu },\beta \}p_{\nu }]  \tag{2.15b}
\end{equation}
\begin{equation}
\{\tilde{x}_{\mu },\tilde{x}_{\nu }\}=(1+\beta )[x_{\mu }\{\beta ,x_{\nu
}\}-x_{\nu }\{\beta ,x_{\mu }\}]  \tag{2.15c}
\end{equation}
written in terms of the old canonical variables. Choosing $\beta =\frac{1}{2}%
p^{2},$ computing the brackets (2.15b) and (2.15c), and finally imposing
constraint (2.13), we arrive at the brackets 
\begin{equation}
\{\tilde{p}_{\mu },\tilde{p}_{\nu }\}=0  \tag{2.16a}
\end{equation}
\begin{equation}
\{\tilde{x}_{\mu },\tilde{p}_{\nu }\}=\delta _{\mu \nu }-p_{\mu }p_{\nu } 
\tag{2.16b}
\end{equation}
\begin{equation}
\{\tilde{x}_{\mu },\tilde{x}_{\nu }\}=-(x_{\mu }p_{\nu }-x_{\nu }p_{\mu }) 
\tag{2.16c}
\end{equation}
while, again imposing constraint (2.13), the transformation equations
(2.14a) and (2.14c) become the identity transformation 
\begin{equation}
x^{\mu }\rightarrow \tilde{x}^{\mu }=x^{\mu }  \tag{2.17a}
\end{equation}
\begin{equation}
p_{\mu }\rightarrow \tilde{p}_{\mu }=p_{\mu }  \tag{2.17b}
\end{equation}
Using equations (2.17) in (2.16), we can then write down the brackets 
\begin{equation}
\{p_{\mu },p_{\nu }\}=0  \tag{2.18a}
\end{equation}
\begin{equation}
\{x_{\mu },p_{\nu }\}=\delta _{\mu \nu }-p_{\mu }p_{\nu }  \tag{2.18b}
\end{equation}
\begin{equation}
\{x_{\mu },x_{\nu }\}=-(x_{\mu }p_{\nu }-x_{\nu }p_{\mu })  \tag{2.18c}
\end{equation}
In the transition to the quantum theory the brackets (2.18) will exactly
reproduce the Snyder commutators [11] originally proposed in 1947 as a way
to solve the ultraviolet divergence problem in quantum field theory.

The term that appears with the minus sign on the right in braket (2.18c) is
the Lorentz generator (2.6b) of rotations in the $d$ dimensional space-time.
It satisfies equation (2.7c) and is gauge invariant because it has vanishing
Poisson bracket with constraint (13), $\{M_{\mu \nu },\phi \}=0.$

\section{Newtonian Gravitodynamics}

The non-relativistic equations for Newtonian gravitodynamics were considered
in [4]. Here we consider the relativistic action 
\begin{equation}
S=\int d\tau (\frac{1}{2}\lambda ^{-1}\dot{x}^{2}+A.\dot{x})  \tag{3.1}
\end{equation}
for a massless particle moving in an external gravitodynamic potential $%
A_{\mu }=A_{\mu }(\tau )$. Action (3.1) should not be interpreted as
describing a massless particle in an electromagnetic background potential
because, as revealed by the ADM [13] construction of general relativity, a
massless neutral particle couples only to the gravitational field.

Action (3.1) is invariant under the infinitesimal reparametrization 
\begin{equation}
\delta x^{\mu }=\epsilon \dot{x}^{\mu }  \tag{3.2a}
\end{equation}
\begin{equation}
\delta \lambda =\frac{d}{d\tau }(\epsilon \lambda )  \tag{3.2b}
\end{equation}
\begin{equation}
\delta A_{\mu }=\epsilon \dot{A}_{\mu }  \tag{3.2c}
\end{equation}
where $\epsilon (\tau )$ is an arbitrary parameter, because the Lagrangian
in action (3.1) transforms as a total derivative, $\delta L=\frac{d}{d\tau }%
(\epsilon L)$, under transformations (3.2). Action (3.1) describes a
reparametrization-invariant system.

Action (3.1) is also invariant under the Poincar\'{e} transformation 
\begin{equation}
\delta x^{\mu }=a^{\mu }+\omega _{\nu }^{\mu }x^{\nu }  \tag{3.3a}
\end{equation}
\begin{equation}
\delta A^{\mu }=b^{\mu }+\omega _{\nu }^{\mu }A^{\nu }  \tag{3.3b}
\end{equation}
\begin{equation}
\delta \lambda =0  \tag{3.3c}
\end{equation}
under the scale transformation 
\begin{equation}
\delta x^{\mu }=\alpha x^{\mu }  \tag{3.4a}
\end{equation}
\begin{equation}
\delta A^{\mu }=-\alpha A^{\mu }  \tag{3.4b}
\end{equation}
\begin{equation}
\delta \lambda =2\alpha \lambda  \tag{3.4c}
\end{equation}
and under the conformal transformation 
\begin{equation}
\delta x^{\mu }=(2x^{\mu }x^{\nu }-\delta ^{\mu \nu }x^{2})b_{\nu } 
\tag{3.5a}
\end{equation}
\begin{equation}
\delta A^{\mu }=2(x^{\mu }A^{\nu }-A^{\mu }x^{\nu }-A.x\delta ^{\mu \nu
})b_{\nu }  \tag{3.5b}
\end{equation}
\begin{equation}
\delta \lambda =4\lambda x.b  \tag{3.5c}
\end{equation}
The gravitodynamic action (3.1) defines a conformal theory in $d$ space-time
dimensions.

The physically relevant canonical momenta that follow from action (3.1) are 
\begin{equation}
p_{\lambda }=0  \tag{3.6}
\end{equation}
\begin{equation}
p_{\mu }=\frac{\dot{x}_{\mu }}{\lambda }+A_{\mu }  \tag{3.7}
\end{equation}
and the corresponding Hamiltonian is 
\begin{equation}
H=\frac{1}{2}\lambda (p_{\mu }-A_{\mu })^{2}  \tag{3.8}
\end{equation}
The Hamiltonian (3.8) tells us that action (3.1) can be obtained by solving
for the momentum in the Lagrangian 
\begin{equation}
L=p.\dot{x}-\frac{1}{2}\lambda (p-A)^{2}  \tag{3.9}
\end{equation}
Lagrangian (3.9) will be our clue to 2T gravity.

Equation (3.6) is a primary constraint. Introducing the Lagrange multiplier $%
\chi (\tau )$ for this constraint, constructing the Dirac Hamiltonian $%
H_{D}=H+\chi p_{\lambda }$ and requiring the dynamical stability of (3.6),
we obtain the constraint 
\begin{equation}
\phi =\frac{1}{2}(p_{\mu }-A_{\mu })^{2}=0  \tag{3.10}
\end{equation}
which is a first-class constraint.

Now, the gravitodynamic Hamiltonian (3.8) is invariant under the
transformations 
\begin{equation}
p_{\mu }\rightarrow \tilde{p}_{\mu }=\exp \{-\beta \}p_{\mu }  \tag{3.11a}
\end{equation}
\begin{equation}
\lambda \rightarrow \exp \{2\beta \}\lambda  \tag{3.11b}
\end{equation}
\begin{equation}
A_{\mu }\rightarrow \tilde{A}_{\mu }=\exp \{-\beta )\}A_{\mu }  \tag{3.11c}
\end{equation}
and from expression (3.7) we find that $x^{\mu }$ must transform as 
\begin{equation}
x^{\mu }\rightarrow \tilde{x}^{\mu }=\exp \{\beta \}x^{\mu }  \tag{3.11d}
\end{equation}
when $p_{\mu }$ transforms as in (3.11a).

To proceed in a consistent way with the construction of the Snyder brackets
for the gravitodynamic action (3.1) it is necessary to introduce, in
addition to brackets (2.8), the new brackets 
\begin{equation}
\{A_{\mu },x_{\nu }\}=x_{\nu }A_{\mu }  \tag{3.12a}
\end{equation}
\begin{equation}
\{A_{\mu },p_{\nu }\}=-p_{\nu }A_{\mu }  \tag{3.12b}
\end{equation}
\begin{equation}
\{A_{\mu },A_{\nu }\}=0  \tag{3.12c}
\end{equation}

The above set of brackets is the $d$ dimensional version of the $d+2$
dimensional brackets introduced in [14].

Choosing now $\beta =\frac{1}{2}(p-A)^{2}$ and performing the same steps as
in the free theory we end with the brackets 
\begin{equation}
\{p_{\mu },p_{\nu }\}=0  \tag{3.13a}
\end{equation}
\begin{equation}
\{x_{\mu },p_{\nu }\}=\delta _{\mu \nu }-(p_{\mu }-A_{\mu })p_{\nu } 
\tag{3.13b}
\end{equation}
\begin{equation}
\{x_{\mu },x_{\nu }\}=-[x_{\mu }(p_{\nu }-A_{\nu })-x_{\nu }(p_{\mu }-A_{\mu
})]  \tag{3.13c}
\end{equation}
which extend the free massless particle Snyder brackets (2.18) to the case
where the massless particle interacts with a background gravitodynamic
potential $A_{\mu }(\tau )$. Notice that the Snyder bracket (3.13c), which
will define the space-time geometry in the quantum theory, is consistent
with the minimal coupling prescription $p_{\mu }\rightarrow p_{\mu }-A_{\mu
} $ applied to the free bracket (2.18c). In other words, the generator
(2.6b) of Lorentz rotations in the $d$ dimensional space-time is modified to 
\begin{equation}
M_{\mu \nu }^{I}=x_{\mu }(p_{\nu }-A_{\nu })-x_{\nu }(p_{\mu }-A_{\mu }) 
\tag{3.14}
\end{equation}
when background gauge fields are introduced. If we use the brackets (2.8)
and (3.12) we will find that $M_{\mu \nu }^{I}$ satisfies an equation
identical to the free equation (2.7c). It also satisfies 
\begin{equation}
\{M_{\mu \nu }^{I},\phi \}=2(x_{\mu }A_{\nu }-x_{\nu }A_{\mu })\phi 
\tag{3.15}
\end{equation}
and is therefore gauge invariant because its bracket with the first-class
constraint (3.10) weakly vanishes, $\{M_{\mu \nu }^{I},\phi \}\approx 0$.

\section{2T Newtonian Gravitodynamics}

Let us now consider how a gravitodynamic model with two times can be
constructed. The first observation in this direction is that action (3.1)
has a $d$ dimensional conformal invariance. As is well known, this
invariance can also be interpreted as the Lorentz invariance in $d+2$
dimensions. To pass to a $d+2$ formalism we must then introduce two extra
first class constraint to appropriately take care of the physical degrees of
freedom. We can then write the 2T gravitodynamic action [14] 
\begin{equation}
S=\int d\tau \{P.\dot{X}-[\frac{1}{2}\lambda _{1}(P-A)^{2}+\lambda
_{2}(P-A).X+\frac{1}{2}\lambda _{3}X^{2}]\}  \tag{4.1}
\end{equation}
where the Hamiltonian is 
\begin{equation}
H=\frac{1}{2}\lambda _{1}(P-A)^{2}+\lambda _{2}(P-A).X+\frac{1}{2}\lambda
_{3}X^{2}  \tag{4.2}
\end{equation}
and the $\lambda $'s are three independent Lagrange multipliers. The
equations of motion for these multipliers now give the primary constraints 
\begin{equation}
\phi _{1}=\frac{1}{2}(P-A)^{2}=0  \tag{4.3}
\end{equation}
\begin{equation}
\phi _{2}=(P-A).X=0  \tag{4.4}
\end{equation}
\begin{equation}
\phi _{3}=\frac{1}{2}X^{2}=0  \tag{4.5}
\end{equation}
To continue it is necessary to introduce the $d+2$ dimensional extension of
brackets (2.8) and (3.12), that is 
\begin{equation}
\{P_{M},P_{N}\}=0  \tag{4.6a}
\end{equation}
\begin{equation}
\{X_{M},P_{N}\}=\delta _{MN}  \tag{4.6b}
\end{equation}
\begin{equation}
\{X_{M},X_{N}\}=0  \tag{4.6c}
\end{equation}
\begin{equation}
\{A_{M},X_{N}\}=X_{N}A_{M}  \tag{4.6d}
\end{equation}
\begin{equation}
\{A_{M},P_{N}\}=-P_{N}A_{M}  \tag{4.6e}
\end{equation}
\begin{equation}
\{A_{M},A_{N}\}=0  \tag{4.6f}
\end{equation}
Computing the algebra of constraints (4.3)-(4.5) using the brackets (4.6),
we obtain the equations 
\begin{equation}
\{\phi _{1},\phi _{2}\}=-2\phi _{1}-2X.A\phi _{1}  \tag{4.7a}
\end{equation}
\begin{equation}
\{\phi _{1},\phi _{2}\}=-\phi _{2}-2(P-A).A\phi _{3}  \tag{4.7b}
\end{equation}
\begin{equation}
\{\phi _{2},\phi _{3}\}=-2\phi _{3}-2X.A\phi _{3}  \tag{4.7c}
\end{equation}
Equations (4.7) show that constraints (4.3)-(4.5) are first-class
constraints. These equations exactly reproduce the $Sp(2,R)$ gauge algebra
of two-time physics when the conditions 
\begin{equation}
X.A=0  \tag{4.8a}
\end{equation}
\begin{equation}
(P-A).A=0  \tag{4.8b}
\end{equation}
hold. Condition (4.8a) is the first of Dirac's subsidiary conditions [15,16]
on the gauge field. Condition (4.8b) is not an independent condition. It is
the condition that is canonically conjugate to condition (4.8a), but
incorporating the minimal coupling prescription $P_{M}\rightarrow
P_{M}-A_{M} $. Conditions (4.8a) and (4.8b) form a single canonical
condition that is necessary for the $Sp(2,R)$ gauge invariance of action
(4.1). When (4.8) holds the only possible metric associated with constraints
(4.3)-(4.5) giving a non-trivial surface and avoiding the ghost problem is
the flat metric with two time-like dimensions ($M=1,...,D+2$).

Hamiltonian (4.2) is now invariant under transformations 
\begin{equation}
X_{M}\rightarrow \tilde{X}_{M}=\exp \{\beta \}X_{M}  \tag{4.9a}
\end{equation}
\begin{equation}
P_{M}\rightarrow \tilde{P}_{M}=\exp \{-\beta \}P_{M}  \tag{4.9b}
\end{equation}
\begin{equation}
A_{M}\rightarrow \tilde{A}_{M}=\exp \{-\beta \}A_{M}  \tag{4.9c}
\end{equation}
\begin{equation}
\lambda _{1}\rightarrow \exp \{2\beta \}\lambda _{1}  \tag{4.9d}
\end{equation}
\begin{equation}
\lambda _{2}\rightarrow \lambda _{2}  \tag{4.9e}
\end{equation}
\begin{equation}
\lambda _{3}\rightarrow \exp \{-2\beta \}\lambda _{3}  \tag{4.9f}
\end{equation}
Choosing $\beta =\frac{1}{2}(P-A)^{2}$ and performing the same steps as in
the previous sections, we obtain the $D+2$ dimensional brackets 
\begin{equation}
\{P_{M},P_{N}\}=0  \tag{4.10a}
\end{equation}
\begin{equation}
\{X_{M},P_{N}\}=\delta _{MN}-(P_{M}-A_{M})P_{N}  \tag{4.10b}
\end{equation}
\begin{equation}
\{X_{M},X_{N}\}=-[X_{M}(P_{N}-A_{N})-X_{N}(P_{M}-A_{M})]  \tag{4.10c}
\end{equation}
From (4.10c) we see that the $SO(d,2)$ generator is 
\begin{equation}
L_{MN}^{I}=X_{M}(P_{N}-A_{N})-X_{N}(P_{M}-A_{M})  \tag{4.11}
\end{equation}
Using brackets (4.6) we find that $L_{MN}^{I}$ satisfies the algebra 
\begin{equation}
\{L_{MN}^{I},L_{RS}^{I}\}=\delta _{MR}L_{NS}^{I}+\delta
_{NS}L_{MR}^{I}-\delta _{MS}L_{NR}^{I}-\delta _{NR}L_{MS}^{I}  \tag{4.12}
\end{equation}
Using again brackets (4.6) we get the equations 
\begin{equation}
\{L_{MN}^{I},\phi _{1}\}=2(X_{M}A_{N}-X_{N}A_{M})\phi _{1}  \tag{4.13a}
\end{equation}
\begin{equation}
\{L_{MN}^{I},\phi _{2}\}=0  \tag{4.13b}
\end{equation}
\begin{equation}
\{L_{MN}^{I},\phi _{3}\}=-2(X_{M}A_{N}-X_{N}A_{M})\phi _{3}  \tag{4.13c}
\end{equation}
which show that $L_{MN}^{I}$ is gauge invariant because it has weakly
vanishing brackets with the first-class constraints (4.3)-(4.5), $%
\{L_{MN}^{I},\phi _{i}\}\approx 0$.

Let us now check the consistency of our formalism. Using Dirac's condition
(4.8a), constraint (4.4) becomes 
\begin{equation}
X.P=X.A=0  \tag{4.14}
\end{equation}
Imposing the gauge conditions [16] $P^{+}(\tau )=A^{+}(\tau )=0$ and $%
X^{+}(\tau )=1,$ and solving constraints (4.14) together with constraint
(4.5), we obtain the expressions [16] 
\begin{equation}
M=(+,-,\mu )  \tag{4.15a}
\end{equation}
\begin{equation}
X^{M}=(1,\frac{x^{2}}{2},x^{\mu })  \tag{4.15b}
\end{equation}
\begin{equation}
P^{M}=(0,x.p,p^{\mu })  \tag{4.15c}
\end{equation}
\begin{equation}
A^{M}=(0,x.A,A^{\mu })  \tag{4.15d}
\end{equation}
In this particular gauge the components of the gauge invariant $L_{MN}^{I}$
become 
\begin{equation}
L_{+\mu }^{I}=p_{\mu }^{I}=p_{\mu }-A_{\mu }  \tag{4.16a}
\end{equation}
\begin{equation}
L_{\mu \nu }^{I}=M_{\mu \nu }^{I}=x_{\mu }(p_{\nu }-A_{\nu })-x_{\nu
}(p_{\mu }-A_{\mu })  \tag{4.16b}
\end{equation}
\begin{equation}
L_{+-}^{I}=D^{I}=x.(p-A)  \tag{4.16c}
\end{equation}
\begin{equation}
L_{\mu -}^{I}=\frac{1}{2}K_{\mu }^{I}=x_{\mu }x.(p-A)-\frac{1}{2}%
x^{2}(p_{\mu }-A_{\mu })  \tag{4.16d}
\end{equation}
The expressions that appear in the right-hand side of equations (4.16) are
respectively the extensions of the free conformal generators (2.6) to the
case when a background gauge field $A_{\mu }$ is present. It can be
verified, using brackets (2.8) and (3.12), that $p_{\mu }^{I}$, $M_{\mu \nu
}^{I}$, $D^{I}$ and $K_{\mu }^{I}$ satisfy the same conformal algebra (2.7)
as in the free theory. These results confirm the consistency of the
relativistic gravitodynamic theory defined by action (3.1).

\section{Concluding remarks}

In this article we tried to give evidence that in the gap between the usual
non-relativistic Newtonian gravitation and the generally covariant
Einsteinian gravitation there may exist another gravitational theory in the
form of a covariant Newtonian gravitational theory.  

We think that some of the features displayed by our model action, such as
reparametrization invariance, conformal invariance, and a consistent
interpretation as a gauge fixed sector of a covariant gravitational theory
with two time-like dimensions, may be considered  as evidences of the
consistency of this covariant Newtonian gravitational theory.

\end{document}